\documentclass[aps,nobibnotes,nofootinbib,a4paper,floatfix,showpacs,preprint,tightenlines,byrevtex,notitlepage]{revtex4}

\newcommand{\est}{{\mathcal{E}}}
\newcommand{\hst}{{\mathcal{H}}}
\newcommand{\qst}{{\mathcal{Q}}}
\newcommand{\vst}{{\mathcal{V}}}
\newcommand{\cst}{{\mathcal{C}}}
\newcommand{\oms}{\omega_1}
\newcommand{\oma}{\omega_2}
\newcommand{\beq}{\begin{equation}}
\newcommand{\eeq}{\end{equation}}
\newcommand{\bear}{\begin{eqnarray}}
\newcommand{\eear}{\end{eqnarray}}

\usepackage{graphicx}

\begin{document}

\preprint{INFN/TC-02/03}

\title{A DETECTOR OF GRAVITATIONAL WAVES BASED ON COUPLED MICROWAVE CAVITIES}

\author{Ph. Bernard} 
\affiliation{CERN, CH--1211, Geneva 23, Switzerland}
\author{A. Chincarini}
\author{G. Gemme}
\thanks{Corresponding author}
\email{gianluca.gemme@ge.infn.it}
\author{R. Parodi}
\affiliation{INFN--Sezione di Genova, via Dodecaneso 33, I--16146 Genova, Italy}
\author{E. Picasso}
\affiliation{Scuola Normale Superiore, Piazza dei Cavalieri 7, I--56126, Pisa, Italy}

\date{March 4, 2002}

\begin{abstract}
Since 1978 superconducting coupled cavities have been proposed as sensitive detector of gravitational waves. The interaction of the gravitational wave with the cavity walls, and the resulting motion, induces the transition of some electromagnetic energy from an initially excited cavity mode to an empty one. The energy transfer is maximum when the frequency of the wave is equal to the frequency difference of the two cavity modes. In this paper the basic principles of the detector are discussed. The interaction of a gravitational wave with the cavity walls is studied in the proper reference frame of the detector, and the coupling between two electromagnetic normal modes induced by the wall motion is analyzed in detail. Noise sources are also considered; in particular the noise coming from the brownian motion of the cavity walls is analyzed. Some ideas for the developement of a realistic detector of gravitational waves are discussed; the outline of a possible detector design and its expected sensitivity are also shown.
\end{abstract}


\maketitle

\section{Introduction}
\label{intro}

In a series of papers it was studied how the effects due to the interaction between the gravitational and the electromagnetic fields could be used to detect gravitational waves \cite{bppr1,bppr2}. The proposed detector exploits the energy transfer induced by the gravitational wave between two levels of an electromagnetic resonator, whose frequencies $\oms$ and $\oma$ are both  much larger than the angular frequency $\Omega$ of the g.w. and satisfy the resonance conditon $|\oma - \oms| = \Omega$.\,\footnote{The interaction between the g.w. and the detector is characterized by a transfer of energy and of angular momentum. Since the elicity of the g.w., i.e. the angular momentum along the direction of propagation, is $2$, it can induce a transition between the two levels provided their angular momenta differ by $2$; this can be achived by putting the two cavities at right angle or by a suitable polarization of the electromagnetic field inside the resonator.} In the scheme suggested by Bernard et al. the two levels are obtained by coupling two identical high frequency cavities\footnote{Throughout this paper, we shall call {\em resonator} the whole detector made up of two coupled {\em cavities}; e.g. we shall speak about one resonator composed by two coupled spherical cavities.}; the angular frequency $\oms$ is the frequency of the level symmetrical in the fields of the two cavities, and $\oma$ is that of the antisymmetrical one. The frequency difference between the symmetric and the antisymmetric level is determined by the coupling, and can be adjusted by a careful resonator design. Since the detector sensitivity is proportional to the square of the resonator quality factor, superconducting cavities should be used for maximum sensitivity.

The power transfer between the levels of a resonator made up of two pill--box cavities, mounted {\em end--to--end} and coupled by a small circular aperture in their common endwall, was checked in a series of experiments by Melissinos et al., where the perturbation of the resonator volume was induced by a piezoelectric crystal \cite{rrm1,rrm2}. Recently the experiment was repeated by our group with an improved experimental set--up; we obtained a sensitivity to fractional deformations of the resonator length as small as $\delta \ell / \ell \approx 10^{-20}$ Hz$^{-1/2}$ \cite{rsi}.

In this paper we shall discuss the mechanism of the interaction of a gravitational wave with a detector based on two coupled resonant cavities. In previous works this issue was discussed using the concept of a dielectric tensor associated with the gravitational wave \cite{pr1}. The interaction was analyzed in the reference frame where the resonator walls were {\em at rest} even in presence of a gravitational perturbation. We shall analyze the effect in the proper reference frame attached to the detector and we shall therefore consider the interaction between the wave and the field stored inside the resonator due to the coupling of the g.w. with the {\em mechanical} structure of the detector \cite{caves}.  

The paper will be organized as follows: in section \ref{sec:elettr} the problem of finding the electromagnetic fields in a closed volume with time--varying boundary conditions is studied, and an approximate expression of the normal modes in a perturbed resonator is worked out. In section \ref{sec:mech} we shall analyze the interaction of a g.w. with the mechanical structure of the detector; we shall see that the transfer of energy between a mechanical and an electromagnetic oscillation depends both on the electromagnetic field distribution inside the resonator and on the resonator geometry and mechanical properties. In section \ref{sec:spherical} we shall discuss some ideas for the developement of a realistic gravitational wave detector based on spherical microwave cavities. Afterwords the coupled equations of motion for the fields in a perturbed resonator are worked out and solved. In section \ref{sec:noise} the issue of the thermal noise of the detector's walls is studied; other noise contribution are also considered. Finally, in the last section, the expected sensitivity of some detector configuration is shown and discussed.

\section{Electromagnetic field in a resonator with perturbed boundaries}
\label{sec:elettr}

To study the mechanism of the energy transfer between the two levels of an electromagnetic resonator perturbed by a gravitational wave we shall follow classic electromagnetic theory. We shall make use of the fact that any field configuration inside the resonator can be expressed as the superposition of the electromagnetic normal modes of the given resonator \cite{slater}.

If no sources are present, the electromagnetic field in vacuum is determined by the equations:
\beq
\label{eq:maxwell1}
\vec\nabla \cdot \vec E = 0
\eeq
\beq
\label{eq:maxwell2}
\vec\nabla \cdot \vec H = 0
\eeq
\beq
\label{eq:maxwell3}
\vec\nabla \wedge \vec E + \mu_0 \frac{\partial \vec H}{\partial t} = 0
\eeq
\beq
\label{eq:maxwell4}
\vec\nabla \wedge \vec H - \epsilon_0 \frac{\partial \vec E}{\partial t} = 0
\eeq
As can be easily verified, eqs. (\ref{eq:maxwell1})--(\ref{eq:maxwell4}) are automatically satisfied if the fields satisfy the wave equations:
\beq
\label{eq:wave1}
\nabla^2 \vec E - \frac{1}{c^2} \frac{\partial^2 \vec E}{\partial t^2} = 0
\eeq
\beq
\label{eq:wave2}
\nabla^2 \vec H - \frac{1}{c^2} \frac{\partial^2 \vec H}{\partial t^2} = 0
\eeq
with $c = (\mu_0\epsilon_0)^{-1/2}$.
 
Let us assume that the field is contained in a resonator with perfectly conducting walls. If we impose boundary conditions on the fields, the solution of the wave equations (\ref{eq:wave1})--(\ref{eq:wave2}) will have an infinite discrete set of normal--mode solutions orthogonal to one another and complete, in the sense that any arbitrary field in the resonator can be expressed as a sum of these normal modes with suitable amplitudes. The amplitudes of each mode can then be used to describe the field in the resonator. 

By the familiar procedure of separation of variables we may assume a solution of eqs. (\ref{eq:wave1})--(\ref{eq:wave2}) of the form:
\beq
\label{eq:expa1}
\vec E(\vec r,t) = \sum_{n=0}^\infty \est_n(t) \vec E_n(\vec r)
\eeq
and 
\beq
\label{eq:expa2}
\vec H(\vec r,t) = \sum_{n=0}^\infty \hst_n(t) \vec H_n(\vec r)
\eeq
where we have defined: 
\beq
\label{eq:estorto}
\est_n(t) \equiv \sqrt{\epsilon_0} \int_V \vec E \cdot \vec E_n \, dV
\eeq 
and
\beq
\label{eq:hstorto} 
\hst_n(t) \equiv \sqrt{\mu_0} \int_V \vec H \cdot \vec H_n \, dV
\eeq 
where the integrals are performed over the resonator volume.

We require that at the walls the tangential component of $\vec E$ and the normal component of $\vec H$ vanish. With this assumption the functions  $\vec E_n(\vec r)$ and $\vec H_n(\vec r)$ satisfy the equations: 
\bear
k_n \vec E_n = \vec\nabla \wedge \vec H_n \\
k_n \vec H_n = \vec\nabla \wedge \vec E_n
\eear
where $k_n=\omega_n/c$ is the propagation constant associated with the $n_{th}$ mode.

It can be proved that the normal modes $\vec E_n$ and $\vec H_n$ have orthogonality properties of the form
\bear
\label{eq:ortho}
\int_V \vec E_n \cdot \vec E_m \, dV & = & \delta_{nm} \\
\int_V \vec H_n \cdot \vec H_m \, dV & = & \delta_{nm}
\eear 

For a cubical resonator the functions $\vec E_n(\vec r)$ and $\vec H_n(\vec r)$ are $\sin(\vec k_n \cdot \vec r)$ and $\cos(\vec k_n \cdot \vec r)$. For other geometries they will be other complete sets of functions. The boundary conditions and geometry determine the different modes which are distinguished by the index $n$. In general three numbers are needed to specify a mode; $n$ is an abbreviation for this set of numbers.

Let us now expand the fields $\vec E$, $\vec H$  in terms of the orthogonal functions $\vec E_n$ and $\vec H_n$; when we substitute these expansions in Maxwell's equations and equate coefficients, so as to get the differential equations satisfied by the various coefficients, we find that equations (\ref{eq:maxwell1}) and (\ref{eq:maxwell2}) are automatically satisfied. From equations (\ref{eq:maxwell3}) and (\ref{eq:maxwell4}) we find the following equations for the expansion coefficients:
\beq
\label{eq:timedep1}
\frac{d \hst_n(t)}{dt} - \omega_n \est_n(t)  = -\frac{\omega_n}{\qst_n}\hst_n(t)
\eeq 
\beq
\label{eq:timedep4}
\frac{d \est_n(t)}{dt} + \omega_n \hst_n(t) = 0 
\eeq

We have taken into account the dissipation arising from the finite conductivity of the walls introducing the electromagnetic quality factor \cite{slater}:
\beq
\label{eq:qdef}
\qst_n = \frac{\omega_n \mu_0}{R_s} \frac{\int_V H_n^2 \, dV}{\int_S H_n^2 \, dS} = \frac{G_n}{R_s} 
\eeq
$R_s$ is the material--dependent surface resistance of the walls, and the geometric factor $G_n$ of the $n_{th}$ mode is defined as:
\beq
\label{eq:gdef}
G_n = \omega_n \mu_0 \frac{\int_V H_n^2 \, dV}{\int_S H_n^2 \, dS} 
\eeq
As can readily be seen, equations (\ref{eq:timedep1})--(\ref{eq:timedep4}) for the field expansion coefficients are decoupled: the modes are independent from one another and behave as simple damped harmonic oscillators. 

\subsection{Perturbation of boundaries}
\label{sec:boundpert}
Let us suppose that the resonator's boundary is perturbed so that the eigenvalues and eigenfunctions of the perturbed resonator differ but little from those of the original one. The perturbation method allows to find the eigenvalues and eigenfunctions of the perturbed problem from the knowledge of the original ones. It basically consists in expanding in power series of a perturbation parameter $\sigma$ the new modes and frequencies \cite{goubau}:
\bear
\label{eq:powser}
\vec E_n'(\vec r) &=& \vec E_n(\vec r) + \sigma \, \vec e_n(\vec r) + {\mathcal{O}}(\sigma^2) \nonumber \\ 
\vec H_n'(\vec r) &=& \vec H_n(\vec r) + \sigma \, \vec h_n(\vec r) + {\mathcal{O}}(\sigma^2) \\ 
k_n' &=& k_n + \sigma \, \kappa_n + {\mathcal{O}}(\sigma^2) \nonumber
\eear

In the perturbed resonator we shall have:
\beq
\label{eq:expa1_new}
\vec E(\vec r,t) = \sum_{n=0}^\infty \tilde\est_n(t) \vec E_n'(\vec r)
\eeq
and 
\beq
\label{eq:expa2_new}
\vec H(\vec r,t) = \sum_{n=0}^\infty \tilde\hst_n(t) \vec H_n'(\vec r)
\eeq
with: 
\beq
\label{eq:estorto_new}
\tilde\est_n(t) \equiv \sqrt{\epsilon_0} \int_{V'} \vec E \cdot \vec E_n' \, dV
\eeq 
and
\beq
\label{eq:hstorto_new} 
\tilde\hst_n(t) \equiv \sqrt{\mu_0} \int_{V'} \vec H \cdot \vec H_n' \, dV
\eeq 
where the integrals are now performed over the perturbed volume $V'$.

The perturbed modes will satisfy the following equations:
\bear
\label{eq:modifeq}
k_n' \vec E_n' = \vec\nabla \wedge \vec H_n' \\
k_n' \vec H_n' = \vec\nabla \wedge \vec E_n'
\eear

When the expansions (\ref{eq:powser}) are inserted into the equations (\ref{eq:modifeq}) we obtain a series of equations determining the various terms of the expansion (\ref{eq:powser}). In the first order approximation, the perturbed fields may be written:
\beq
\label{eq:epert} 
\sigma \, \vec e_n(\vec r) = \sum_{m=1}^\infty A_{nm}\vec E_m(\vec r)
\eeq
\beq
\label{eq:hpert} 
\sigma \, \vec h_n(\vec r) = -\frac{1}{2}\cst_{nn}\vec H_n(\vec r) + \sum_{m=1}^\infty B_{nm}\vec H_m(\vec r)
\eeq
\beq
\label{eq:expcoeffk}
\sigma \, \kappa_n = -\frac{1}{2} k_n\,\cst_{nn}
\eeq
where the sums have to be performed for $n \neq m$.
The expansion coefficients have the form \cite{goubau}:
\beq
\label{eq:expcoeffa}
A_{nm} = \frac{k_m k_n}{k_m^2 - k_n^2}\,\cst_{nm}
\eeq
\beq
\label{eq:expcoeffb}
B_{nm} = \frac{k_n^2}{k_m^2 - k_n^2}\,\cst_{nm}
\eeq
with
\beq
\label{eq:cnm}
\cst_{nm} = \int_{\vst} (\vec {H}_n \cdot \vec {H}_m - \vec {E}_n \cdot \vec {E}_m) \, dV
\eeq 
where $\vst = V' - V$ is the (algebraic) difference between the perturbed and the original volume.\footnote{The derivation of the eigenmodes and eigenvalues of the perturbed resonator has been made assuming a {\em static} perturbation; in the following we shall assume that it is also correct for a perturbation which has a rate of change much slower than the e.m. field characteristic frequency.}

The calculation of the coupling coefficient $\cst_{nm}$ depends on how the resonator is deformed by an external force. For this reason in the next section we shall briefly review the study of the mechanical behaviour of a body under the influence of an external force.

\section{Analysis of the mechanical response of the detector}
\label{sec:mech}
The interaction of a gravitational wave with the mechanical structure of the detector can be studied by means of classical, non--relativistic, linear elasticity theory \cite{landau_elas,lobo}.
If $\vec u(\vec r)$ denotes the displacement of the mass element at point $\vec r$, relative to the centre of mass of the body in its unperturbed state, and $\vec f(\vec r,t)$ is the volume force density which acts on the body, the displacement is the solution of the system of partial differential equations:
\beq
\label{eq:elastic}
\rho \frac{\partial^2 \vec u}{\partial t^2}- \mu \nabla^2\vec u - (\lambda + \mu) \vec\nabla (\vec\nabla \cdot \vec u ) = \vec f(\vec r,t)
\eeq
with suitable boundary and initial conditions. $\rho(\vec r)$ is the mass density of the body and $\lambda$ and $\mu$ are the material's elastic Lam\'e coefficients. In the following we shall adopt null initial conditions:
\beq
\label{eq:inicons}
\vec u (\vec r,0) = \frac{\partial\vec u}{\partial t}(\vec r,0) = 0
\eeq

The expansion theorem \cite{saulson} states that the displacement of a system in response to an applied force is equal to the superposition of the normal modes $\vec\xi_\alpha(\vec r)$ of the system\footnote{In this section we shall label with greek indices the {\em mechanical} normal modes of the system, and with latin indices the normal modes of the {\em electromagnetic field} stored inside the system.}:
\beq
\label{eq:expansion}
\vec u(\vec r,t) = \sum_{\alpha=1}^\infty \vec\xi_\alpha(\vec r) q_\alpha(t)
\eeq
The normal modes $\vec \xi_{\alpha}(\vec r)$ are the eigen--solutions to
\beq
\label{eq:eigeneq}
\mu \nabla^2\vec \xi_{\alpha} + (\lambda + \mu) \vec\nabla (\vec\nabla \cdot \vec \xi_{\alpha} ) = -\omega_{\alpha}^2 \rho \vec \xi_{\alpha}
\eeq
with boundary conditions; here $\alpha$ is an index, or set of indices, labelling the mode of frequency $\omega_{\alpha}$.
The modes are normalized so that
\beq
\label{eq:norm}
\int_{Vol} \vec\xi_\alpha(\vec r) \cdot \vec\xi_\beta(\vec r)  \rho(\vec r) \, dV = M_\alpha\delta_{\alpha\beta}
\eeq 
where $M_\alpha$ is the reduced mass of the $\alpha$ mode. For a homogeneous system $M_\alpha \equiv M$, where $M$ is
the mass of the system.

$q_\alpha(t)$ is the generalized coordinate of the $\alpha$ mode, obeying the dynamical equation of motion:
\beq
\label{eq:modeeq}
\ddot q_\alpha(t) + \frac{\omega_\alpha}{Q_\alpha}\,\dot q_\alpha(t) + \omega_\alpha^2 q_\alpha(t) = \frac{f_\alpha(t)}{M} 
\eeq
where an empirical damping term, proportional to the system velocity, has been added; $f_\alpha(t)$ is the generalized force, given by
\beq
\label{eq:genfor}
f_\alpha(t) = \int_{Vol} \vec f(\vec r,t) \cdot \vec\xi_\alpha(\vec r) \, dV
\eeq

The solution of equation (\ref{eq:modeeq}), satisfying the initial conditions (\ref{eq:inicons}), can be written in term of a Green function integral as \cite{courant}:
\beq
\label{eq:qdit}
q_{\alpha}(t) = \frac{1}{M \omega_{\alpha}} \int_0^t f_{\alpha}(t') \sin{\omega_{\alpha}(t-t')}\exp{-\frac{t-t'}{\tau_{\alpha}}}\, dt'
\eeq
with $\tau_{\alpha} = 2Q_{\alpha}/\omega_{\alpha}$ is the amplitude decay time of the system.

In the frequency domain the asymptotic solution of equation (\ref{eq:modeeq}) is easily found to be:
\beq
\label{eq:qomega}
q_\alpha(\omega) = \frac{f_\alpha(\omega)/M}{\omega_\alpha^2 - \omega^2 +j \, \omega \omega_\alpha/Q_\alpha}
\eeq
Substituting into eq. (\ref{eq:expansion}), we find
\beq
\label{eq:expomega}
\vec u(\vec r, \omega) = \sum_{\alpha=1}^\infty \frac{f_\alpha(\omega) \vec\xi_\alpha(\vec r) / M}{\omega_\alpha^2 - \omega^2 +j \, \omega \omega_\alpha/Q_\alpha}
\eeq
It is clear from eq. (\ref{eq:expomega}), that if we are interested in the displacement in a narrow frequency interval $\omega \pm \delta\omega$, only those modes for which $\omega_\alpha \approx \omega$ (and $f_\alpha \neq 0$), will give a significant contribution.

\subsection{Interaction of a g.w. with the mechanical structure of the detector}
\label{sec:gwmech}
An incoming gravitational wave manifests itself as a tidal force density acting on the mechanical structure of the detector. Given the expression of the gravitational force and the mechanical properties of the detector, the resulting deformation can be calculated, with the aid of the mathematical apparatus outlined in the previous section.

\begin{figure}[p]
\includegraphics{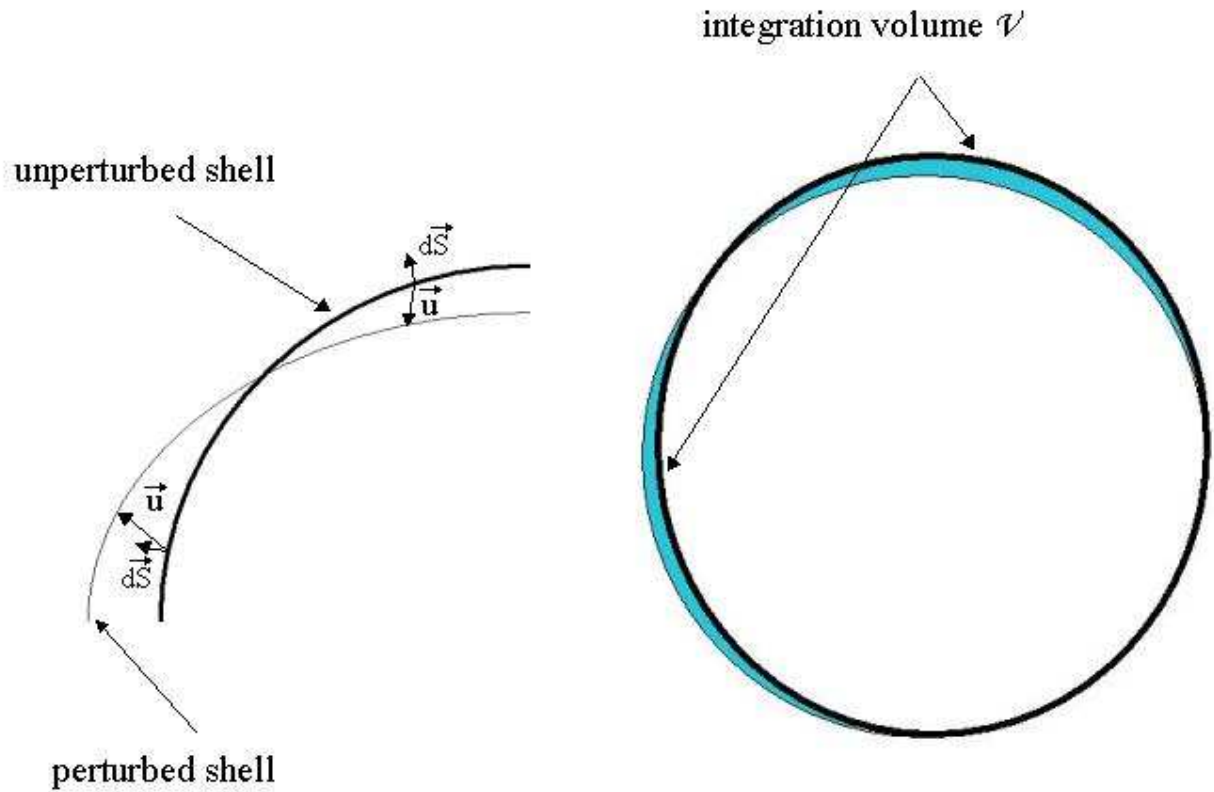}
\caption{\label{fig:deforvol} 
Schematic view of the deformed boundary. $\vec u$ is the local displacement vector; $d\vec S$ is a vector pointing in the direction of the {\em outer} normal of the original surface.} 
\end{figure}

We are mainly interested in the evaluation of the coupling coefficient $\cst_{ij}(t)$. Let us note that, for small displacements, we can write the integral over the perturbed volume as a surface integral in the form (see fig. \ref{fig:deforvol}):
\bear
\label{eq:surfintp}
\lefteqn{\cst_{ij}(t) = \int_{\vst} (\vec {H}_i \cdot \vec {H}_j - \vec {E}_i \cdot \vec {E}_j) \, dV \approx}\nonumber \\ 
& & \int_{S} (\vec {H}_i \cdot \vec {H}_j - \vec {E}_i \cdot \vec {E}_j) \, \vec u(t) \cdot d\vec S
\eear  
where the integral in the r.h.s of eq. (\ref{eq:surfintp}) is now performed over the {\em unperturbed} detector boundary. It is worth noting that this integral can be expressed as a superposition of the mechanical normal modes of the system. Using the expansion theorem (\ref{eq:expansion}) we can write:
\bear
\label{eq:vtilde}
\lefteqn{\cst_{ij}(t) = \int_{S} (\vec {H}_i \cdot \vec {H}_j - \vec {E}_i \cdot \vec {E}_j) \, \vec u(t) \cdot d\vec S =} \nonumber \\
& & \sum_{\alpha=1}^{\infty} q_{\alpha}(t) \int_{S} ( \vec {H}_i \cdot \vec {H}_j - \vec {E}_i \cdot \vec E_j) \, \vec \xi_{\alpha} \cdot d\vec S =
\sum_{\alpha=1}^\infty q_{\alpha}(t) C^{\alpha}_{ij}
\eear
where we have defined the time--independent form factor $C^{\alpha}_{ij}$ as:\footnote{We remind that the superscript $\alpha$ labels the {\em mechanical} normal mode, while the subscripts $i$ and $j$ label the {\em electromagnetic} modes.}
\beq
\label{eq:gammap}
C^{\alpha}_{ij} = \int_{{S}} (\vec {H}_i \cdot \vec {H}_j - \vec {E}_i \cdot \vec {E}_j) \, \vec \xi_{\alpha} \cdot d\vec S
\eeq
If the external force couples strongly only to one mechanical mode of the detector (say the $m$), we can write the simplified expression
\beq
\label{eq:gammapsimple}
\cst_{ij}(t) = q_m(t) C^{m}_{ij}
\eeq

In summary, to have an effective coupling between the two electromagnetic modes we need:
\begin{enumerate}
\item{that the generalized coordinate $q_m(t)$ is different from zero. If the system is initially at rest, this is true if and only if the generalized force, is itself different from zero, as is shown in eqs. (\ref{eq:modeeq}) and (\ref{eq:qdit})};
\item{that the spatial integral in eq. (\ref{eq:gammap}) is different from zero. In section \ref{sec:rect} we shall discuss 
how this depends on the symmetries of the electromagnetic field {\em and} of the perturbed volume.}
\end{enumerate}

\section{Detector design}
\label{sec:spherical}
\begin{figure}[p]
\includegraphics{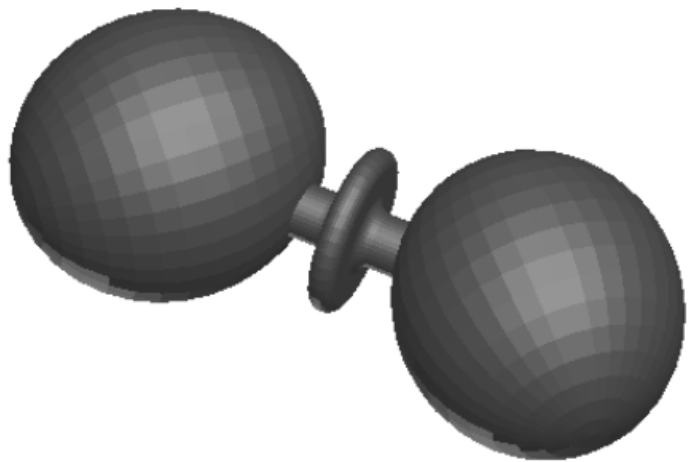}
\caption{\label{fig:paco2} 
Artistic view of the coupled spherical cavities with the central tuning cell.} 
\end{figure}

In order to build a realistic detector a suitable cavity shape has to be chosen. From quite general arguments a detector based on two coupled spherical cavites looks very promising (see fig. \ref{fig:paco2}).

In order to approach the 
interesting frequency range for g.w. detection, the mode splitting (i.e. the detection 
frequency) will be $\oma - \oms \approx 10$ kHz. The internal radius of the spherical cavity will be $r \approx 
100$ mm, corresponding to a frequency of the TE$_{011}$ mode $\omega \approx 2$ GHz. The overall system mass and length
will be $M \approx 5$ kg and $L \approx 0.8$ m. The choice of these frequencies for the resonator and mode splitting
will be also useful in order to test the feasibility of a detector working at $\approx 200$ MHz and at a detection frequency
of $\approx 1$ KHz. 

A tuning cell, or a 
superconducting bellow, will be inserted in the coupling tube between the two cavities, 
allowing to tune the coupling strength (i.e. the detection frequency) in a narrow range 
around the design value. 

From the point of view of the electromagnetic design the spherical cell has the 
highest geometrical factor, and so the highest quality factor, for a given surface 
resistance. For the TE$_{011}$ mode of a sphere the geometric factor $G$ has a value $G \approx 850 \, \Omega$, while 
for a standard elliptical accelerating cavity the TM$_{010}$ mode has a value of $G \approx 250 \, \Omega$. 
Looking at the best reported values of quality factor of accelerating cavities, which 
typically are in the range $10^{10}$--$10^{11}$, we can extrapolate that the quality factor of the TE$_{011}$ 
mode of a spherical cavity can exceed ${\cal{Q}} \approx 10^{11}$.

From the mechanical point of view it is well know that a sphere has the highest 
interaction cross-section with a g.w. and that only a few mechanical modes of the 
sphere do interact with a gravitational perturbation (the quadrupolar ones) \cite{lobo}.
The mechanical design is highly simplified if the spherical geometry is used since the 
deformation of the sphere is given by the superposition of just one or two normal modes of 
vibration and thus can be easily modeled. In fact the proposed detector acts essentially as a 
standard g.w. resonant bar detector: the gravitational perturbation interacts with the 
mechanical structure of the resonator, deforming it. The e.m. field stored inside the 
resonator is affected by the time--varying boundary conditions and a small quantity of 
energy is transferred from the initially excited e.m. mode to the initially empty one, 
provided the g.w. frequency equals the frequency difference of the two modes. We emphasize that our detector is sensitive to the polarization of the incoming gravitational signal: once the e.m. axis has been chosen inside the resonator, a g.w with polarization axes in the direction of the field axis will drive the energy transfer between the two modes of the cavity with maximum efficiency. 

\begin{figure}[p]
\includegraphics{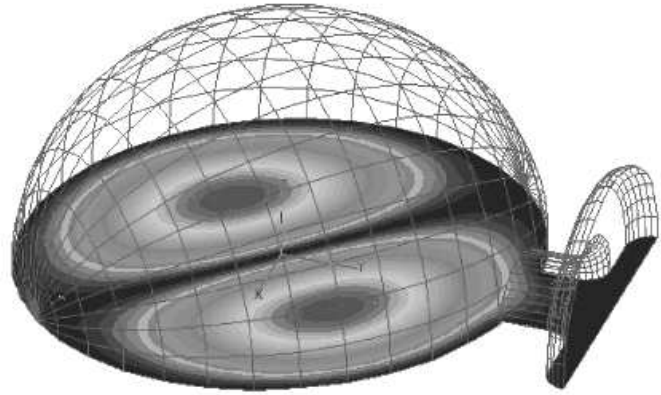}
\caption{\label{fig:magexy} 
Electric field magnitude of the TE$_{011}$ mode. Note the alignment of the field axis.} 
\end{figure}

Finally the spherical cells can be esily deformed in order to remove the unwanted e.m. 
modes degeneracy and to induce the field polarization suitable for g.w. detection.
The interaction between the stored e.m. field and the time-varying boundary 
conditions is not trivial and depends both on how the boundary is deformed by the external 
perturbation and on the spatial distribution of the fields inside the resonator, as shown by the expression of the coupling coefficient $C_{21}^m$ (eq. (\ref{eq:gammap})). It has been 
calculated that the optimal field spatial distribution is with the field axis of the two cavities 
orthogonal to each other (see fig. \ref{fig:magexy}). Different spatial distributions (e.g. with the field axis along the resonators' axis) give a smaller effect or no effect at all. A more detailed discussion of the coupling coefficient calculation is done in section \ref{sec:rect}.

\section{Field equations}
\label{sec:ba}
We shall derive the equations of motion for the fields in the coupled system from a general hamiltonian formalism \cite{louirad}. The hamiltonian of the electromagnetic field inside a resonator with perfectly conducting walls can be written, in terms of the fields amplitudes:
\beq
\label{eq:freeham}
H = \frac{1}{2}\int_V \left( \epsilon_0 \vec E \cdot \vec E + \mu_0 \vec H \cdot \vec H \right ) \, dV
\eeq
If we substitute eqs. (\ref{eq:expa1})--(\ref{eq:expa2}) in eq. (\ref{eq:freeham}) and use the orthonormality condition, the hamiltonian for the field becomes:
\beq
\label{eq:hammodi}
H = \frac{1}{2}\sum_n \left( \est^2_n +\hst_n^2\right ) 
\eeq
If we define a generalized coordinate $X_n$ as:
\beq
\label{eq:qgen}
X_n = \frac{\hst_n}{\omega_n}
\eeq
and its conjugate momentum as:
\beq
\label{eq:pgen}
P_n = \est_n
\eeq
the hamiltonian can be written as:
\beq
\label{eq:hamgen}
H = \frac{1}{2}\sum_n \left( P_n^2 + \omega_n^2 X_n^2\right ) 
\eeq
which is identical to the hamiltonian of an infinite set of uncoupled harmonic oscillators.
It can be easily verified that the field equations of motion can be derived from this hamiltonian by:
\bear
\label{eq:hamilteq}
\frac{\partial H}{\partial X_n} &=& - \dot P_n = \omega^2_n X_n \nonumber \\
\frac{\partial H}{\partial P_n} &=&  \dot X_n = P_n
\eear
which, in terms of the fields become:
\bear
\label{eq:hamilteqf}
\frac{d \est_n}{d t} &=& -\omega_n \hst_n \nonumber \\
\frac{d \hst_n}{d t} &=& \omega_n \est_n
\eear
which are identical to eqs. (\ref{eq:timedep1})--(\ref{eq:timedep4}) if wall dissipation is neglected.

\subsection{Field equations for the perturbed system}
\label{sec:pertequa}
In order to find the equations of motions for the fields in the perturbed resonator we shall make use of the results obtained in the previous sections. Let us consider again an external time--dependent perturbation, whose rate of change is much less than the rate of change of the fields. The normal modes in the perturbed resonator are given by eqs. (\ref{eq:powser})--(\ref{eq:cnm}). If we are looking for the fields in a frequency region where only two electromagnetic modes give a significant contribution, the hamiltonian of the system, as a function of the {\em perturbed} modes amplitudes (see eqs. (\ref{eq:estorto_new}) and (\ref{eq:hstorto_new}), will be:
\beq
\label{eq:hamilpert}
H = \frac{1}{2} \left ( \tilde \est_1^2 + \tilde \hst_1^2 + \tilde \est_2^2 +\tilde \hst_2^2 \right )
\eeq
If we now substitute in the above expression the expansions (\ref{eq:powser}) we obtain the hamiltonian written in terms of the {\em unperturbed} modes amplitudes:
\bear
\label{eq:hampert2}
H = \frac{1}{2} \left (\est_1^2 + \hst_1^2 + \est_2^2 + \hst_2^2 + \frac{p_m^2}{M} + M\omega_m^2 q_m^2 \right ) 
+ \nonumber \\ \frac{q_m}{2} \left ( C_{11}^m \hst_1^2 + C_{22}^m \hst_2^2 + 2\,C_{12}^m \hst_1\hst_2 \right ) -q_m\,f_m
\eear
where the hamiltonian of a mechanical harmonic oscillator, coupled to an external, time--dependent force, has been included.

From this hamiltonian the equations of motion for the fields can readily be obtained:
\beq
\label{eq:fullsys1}
\frac{d \hst_1}{dt} - \omega_1 \est_1 = -\frac{\omega_1}{\qst_1}\hst_1
\eeq
\beq
\label{eq:fullsys2}
\frac{d \hst_2}{dt} - \omega_2 \est_2= -\frac{\omega_2}{\qst_2}\hst_2
\eeq
\beq
\label{eq:fullsys3}
\frac{d \est_1}{dt} + \omega_1 \hst_1 = -\oms q_{m} \left ( C^m_{11} \hst_1 + C^m_{12} \hst_2 \right )
\eeq
\beq
\label{eq:fullsys4}
\frac{d \est_2}{dt} + \omega_2 \hst_2 = -\oma q_{m} \left ( C^m_{12} \hst_1 + C^m_{22}\hst_2 \right )
\eeq
\beq
\frac{d q_m}{dt} - \frac{p_m}{M} = -\frac{\omega_m}{Q_m}q_m
\eeq
\beq
\label{eq:fullsys5}
\frac{d p_m}{dt} + M \omega_m^2 q_m = f_{m} - f^{ba}_{m}
\eeq
where the dissipative terms have been added {\em by hand} and where the term $f^{ba}_{m}$, which describes the back--action effect of the fields on the walls is given by:
\beq
\label{eq:fba}
f^{ba}_{m} = \frac{1}{2} \left( C_{11}^m\hst_1^2 + C_{22}^m\hst_2^2 \right ) + C_{12}\hst_1\hst_2
\eeq

\section{Coupling coefficient calculation}
\label{sec:rect}
The explicit calculation of the coupling coefficient $C_{21}^m$ is not trivial for an arbitrary deformation of the resonator volume and in general can be done only by numerical methods. 

First let us note a general property of the coupling coefficients. Let us consider the resonant modes of the two coupled cavities. As previously noted we have the symmetric and the antisymmetric mode; for the former the electric field $\vec E$ and the magnetic field $\vec H$ in the first cavity are equal to the electric and magnetic field in the second cavity, while for the latter $\vec E$ and $\vec H$ in the first cavity are equal respectively to $- \vec E$ and $-\vec H$ in the second one. From this it follows that in the definition of $C_{21}$, the integrand expression $(\vec H_2 \cdot \vec H_1 - \vec E_2 \cdot \vec E_1)$ -- where we remind that the subscript $1$ indicates the symmetric mode and $2$ the antisymmetric mode -- is {\em odd} over the whole detector volume. For this reason if the volume perturbation, over which we perform the integration, is symmetric between the two cavities, the coupling coefficient vanishes, because the contributions to the integral coming from the two cavities, cancel each other. Otherwise, if the volume perturbation is antisymmetric (when one cavity shrinks, the other expands) the two contributions are added with the same sign, and the coupling coefficient is maximum. This general property suggests that we must find a geometrical configuration of our detector such that the volume deformation due to a g.w. is antisymmetric for the two cavities. This was pointed out already in previous works, where the argument was based on the fact that, since the g.w. carries an angular momentum equal to $2$, the angular momenta of the fields of the two modes should differ by $2$. This can be achieved by putting the two cavities at right angle or by a suitable polarization of the electromagnetic field inside the resonator.

These concepts were verified by both analytical and numerical calculations. General arguments suggested that for an ideal spherical hollow resonator, excited in the fundamental quadrupolar mechanical mode and in the TE$_{011}$ electromagnetic mode, we should have\footnote{We remind that for a TE e.m. mode we have vanishing electric field on the resonator's surface; for this reason the electric field plays no role in the coupling coefficient calculation.}:
\bear
{C_{21}^m} = {C_{12}^m} & = & \int_S \left ( \vec H_1 \cdot \vec H_2 \right ) \vec \xi_m \cdot d\vec S = 0.4 \nonumber\\
{C_{11}^m} & = & \int_S \left ( \vec H_1 \cdot \vec H_1 \right ) \vec \xi_m \cdot d\vec S = 0 \\
{C_{22}^m} & = & \int_S \left ( \vec H_2 \cdot \vec H_2 \right ) \vec \xi_m \cdot d\vec S = 0 \nonumber
\eear

More detailed calculations, made on a realistic model of the coupled spheres, including the central coupling cell and the e.m. input and output ports, were made by finite element methods. These calculations showed that $C_{11}^m = C_{11}^m \leq 10^{-4}$, while $C_{12}^m = C_{21}^m \approx 0.2$.

\section{Calculation of the detector's signal}
\label{sec:sigcalc}
As already pointed out in the introduction, the proposed detector exploits the energy transfer induced by the gravitational wave between two levels of an electromagnetic resonator, whose frequencies $\oms$ and $\oma$ are both  much larger than the angular frequency $\Omega$ of the g.w. and satisfy the resonance conditon $\oma - \oms \approx \Omega$. This is an example of a frequency converter, i.e. a nonlinear device in which energy is transferred from a reference frequency to a different frequency by  an external pump signal. This can be viewed as a three-bodies interaction (given by the field--wall interaction term in the hamiltonian (\ref{eq:hampert2})) which corresponds to annihilation of quanta at $\oms$ and $\omega_m$ and creation at $\oma$ (or vice versa).
For this reason we could argue that, since for a small perturbation $\hst_1(t)$ and $\est_1(t)$ will approximately be sinusoidal functions at frequency $\oms$, while $\hst_2(t)$ and $\est_2(t)$ will oscillate at frequency $\oma$, in the hamiltonian only the  terms varying as $\omega_2 - \omega_1$ will give a significant contribution to the interaction.
The d--c terms will just give an average deformation of the detector's walls, determining a static frequency shift of the resonant modes, while the rapidly fluctuating terms at $\oma + \oms$ would practically average to zero. 

We shall now calculate the field that is excited by the boundary pertubation in mode $2$, starting from an initial condition with mode $1$ strongly excited in the resonator.
To simplify the analysis of the system of differential equations (\ref{eq:fullsys1})--(\ref{eq:fullsys5}) we will
neglect the small perturbation, due to the external force, on the initially excited e.m. mode (mode 1), and will set $\hst_1 \approx A_1 \cos(\oms t)$ and $\est_1 \approx A_1 \sin(\oms t)$, with {\em constant} amplitude $A_1$.\footnote{Actually the amplitude of mode $1$ is kept constant by an external rf power source.} 
Furthermore we shall consider the coupling between two TE modes of a resonator: 
for these modes we have vanishing electric field on the resonator surface. Switching to the complex notation,\footnote{In the following it is understood that the physical fields are the real parts of the complex quantities.} we obtain:
\beq
f^{ba}_{m} \approx \frac{1}{2} C^m_{21}\, \Re(\hst_2\hst_1^*)
\eeq
Finally, to further simplify our calculations, we shall choose a resonator geometry and e.m. field distribution so that 
$C^{m}_{11} = C^{m}_{22} = 0$. We shall see in a following section that this choice is always possible.

With this assumptions, and taking $\qst_1 \approx \qst_2 \equiv \qst$, we can recast the coupled system of equations in the following form:
\beq
\label{eq:simsys1}
\ddot{\cal H}_2 + \frac{\omega_2}{\mathcal{Q}} \dot{\cal H}_2 + \omega_2^2 {\cal H}_2= 
-\omega_2^2 q_m C^m_{21} {\cal H}_1  
\eeq
\beq
\label{eq:simsys2}
\ddot q_m + \frac{\omega_m}{Q_m}\,\dot q_m + \omega_m^2 q_m = \frac{f_{m}}{M} -
\frac{1}{2}\frac{C^{m}_{21}}{M} \, \hst_2\hst_1^*
\eeq

We apply the following substitutions:
\bear
\label{eq:subst}
\hst_2(t) &=& A_2(t)\exp(i \,\oma t) \nonumber\\
q_m(t) &=& Q(t)\exp(i\, \Omega t) \\
f_m(t) &=& F(t)\exp(i\, \Omega t)\nonumber
\eear

Eqs. (\ref{eq:simsys1})--(\ref{eq:simsys2}) now become:
\bear
\label{eq:redusys}
\ddot A_2 + a \dot A_2 + b A_2 &=& c \, Q \nonumber \\
\ddot Q + d \dot Q + e Q &=& g A_2 + F/M
\eear
where $a$, $b$, $c$, $d$, $e$ and $g$ are constant coefficients defined by
\bear
\label{eq:coeffs}
a &=& 2i\,\left(\oms+\Omega\right)+\frac{\oma}{\qst}\nonumber \\
b &=& \oma^2 - \left( \oms+\Omega \right)^2+i\,\frac{\oma}{\qst}\left( \oms+\Omega \right)\nonumber \\
c &=& -\oma^2 C^m_{21} A_1\nonumber \\
d &=& 2i\,\Omega + \frac{\omega_m}{Q_m} \\
e &=& \omega_m^2 - \Omega^2 + i\,\frac{\omega_m \Omega}{Q_m}\nonumber \\
g &=& -\frac{1}{2}\frac{C^m_{21} A_1}{M}\nonumber \\
\eear

We can now fourier transform eqs. (\ref{eq:redusys}) and solve them for $A_2(\omega)$. We find:
\beq
\label{eq:solua2}
A_2(\omega) = -2 \pi \frac{c \, F(\omega)/M}{\left(b-\omega^2+i\,a \omega\right)\left(e-\omega^2 + i\, d \omega\right) - g c}
\eeq

$\hst_2(t)$ is then given by:
\beq
\hst_2(t) = -\exp(i\,\oma t) \int_{-\infty}^{\infty} \frac{c \, F(\omega)/M \,\exp(i\,\omega t)}{\left(b-\omega^2+i\,a \omega\right)\left(e-\omega^2 + i\, d \omega\right) - g c} \, d\omega
\eeq

For a plane g.w travelling along the $z$ axis the force density, in the proper reference frame attached to the detector, 
has the form:
\begin{equation}
\label{eq:gwfor}
\vec f(\vec x,t) = -\frac{1}{2} \rho(\vec x) \left 
[ (\ddot a^1_{\,1} x + \ddot a^1_{\,2} y), (\ddot a^1_{\,2} x - \ddot a^1_{\,1} y), (0)\right ]
\end{equation}
where $a^i_j(t)$, is the adimensional amplitude of the wave, and $a^1_{\,1} = - a^2_{\,2}$, $a^1_{\,2} = a^2_{\,1}$.
The generalized force, acting on the $m$ mechanical mode, then has the form
\bear
\label{eq:gengwfor}
f_m = -\frac{1}{2} \,\ddot a^1_{\,1} \int_{Vol} \left((\xi_m)_x \, x - (\xi_m)_y \, y \right) \rho(\vec x) \, dV - \nonumber \\ \frac{1}{2}\,\ddot a^1_{\,2} \int_{Vol} \left((\xi_m)_x \, y + (\xi_m)_y \, x \right ) \rho(\vec x) \, dV 
\eear
If $a^i_{\,j}(t)$ is given by:
\beq
a^i_{\,j}(t) = \left ( \begin{array}{cc} h^1_{\,1}\, \alpha(t) & h^1_{\,2}\, \beta(t) \\
h^1_{\,2}\, \beta(t) & -h^1_{\,1}\, \alpha(t) \end{array} \right )
\eeq
where $\alpha(t)$ and $\beta(t)$ are sinusoidal functions of frequency $\Omega/(2\pi)$,
then $\ddot a^i_j = -\Omega^2 a^i_j$, and if we define the effective lengths of our detector as 
\bear
\label{eq:efflen}
L_+ &=& \frac{1}{M}\int_{Vol} \left((\xi_m)_x \, x - (\xi_m)_y \, y \right) \rho(\vec x) \, dV \nonumber \\
L_\times &=& \frac{1}{M}\int_{Vol} \left((\xi_m)_x \, y +(\xi_m)_y \, x \right ) \rho(\vec x) \, dV 
\eear
we can write
\beq
f_m(t) = \frac{1}{2} M \Omega^2 \left ( L_+ a^1_{\,1} + L_\times a^1_{\,2} \right ) 
\eeq
and
\beq
A_2(\omega) = -\frac{1}{2}\frac{c \,\Omega^2 \left ( L_+ h^1_{\,1} + L_\times h^1_{\,2} \right )}{b e - g c} \,\delta(\omega - \Omega)
\eeq
or, making use of eq. (\ref{eq:coeffs})
\beq
\hst_2(t) = \frac{1}{2}
\frac{\oma^2 C^m_{21} A_1 \left ( L_+ h^1_{\,1} + L_\times h^1_{\,2} \right ) \Omega^2 \exp(i\,\oma t)}{\left( \omega_m^2-\Omega^2 +i\,\frac{\omega_m\Omega}{Q_m} \right) 
\left( \oma^2-(\oms+\Omega)^2 +i\,\frac{\oma(\oms+\Omega)}{\qst} \right) - \frac{(\oma C^m_{21} A_1)^2}{2M}}
\eeq

The electric field amplitude the initially empty mode can be readily obtained from eq. (\ref{eq:fullsys2}); the average energy stored in mode number 2 is given by: $U_2 = (1/2) |\hst_2|^2 = (1/2) |\est_2|^2$.

\section{Noise issues}
\label{sec:noise}

\subsection{Mechanical thermal noise}
Thermal noise is one of the fundamental limits in the measurement of small displacements. In particular it is one of the dominant noise sources in resonant--mass detectors of g.w. and a major reason that such detectors operate at cryogenic temperatures. Since our detector exploits the coupling of the g.w. with the mechanical structure of the resonator, we have to carefully study the thermal noise contribution to our output signal. 

We start again from eqs. (\ref{eq:simsys1})--(\ref{eq:simsys2}) taking now the external force $f_m(t)$ as a stochastic force with constant power spectrum $S_{ff}$, given by \cite{papoulis}:
\beq
\label{eq:powspec}
S_{ff} = \frac{4Mk_BT\omega_m}{Q_m}
\eeq

Making the substitutions:
\bear
\label{eq:newsubst}
\hst_2(t) &=& A_2(t)\exp(i\, \oma t) \nonumber \\
q_m(t) &=& Q(t)\exp(i\, (\oma-\oms) t)
\eear
we obtain the following equations:
\bear
\label{eq:newredusys}
\ddot A_2 + a \dot A_2 + b A_2 &=& c \, Q \nonumber\\ 
\ddot Q + d \dot Q + e Q &=& g A_2 +  \frac{f(t)}{M} \exp(i\, (\oms - \oma) t)
\eear
where $a$, $b$, $c$, $d$, $e$ and $g$ are defined as in eq. (\ref{eq:coeffs}) with the parameter $\Omega$ replaced by the difference $\oma-\oms$.

The first equation in (\ref{eq:newredusys}) can be solved for $Q(t)$, and we are left with one equation for the variable $A_2(t)$:
\bear
-\frac{d^4 A_2}{d t^4}-(d+a)\frac{d^3 A_2}{d t^3}-(b+e+ad)\frac{d^2 A_2}{d t^2}
-(db+ea)\frac{d A_2}{d t}+(cg-eb)A_2  = \nonumber \\  c\,\frac{f(t)}{M} \exp(i\, (\oms-\oma) t)
\eear

For a linear system we can immediately write the spectral density of the amplitude $A_2$ as \cite{papoulis}:
\beq
S_{AA}(\omega) = | p(\omega) |^{-2} \left(\frac{c}{M}\right)^2 S_{ff}
\eeq
with
\beq
p(\omega) = -\omega^4 + i\,(a+d) \omega^3 + (e + b + da) \omega^2 - i\,(ea + db) \omega +  cg - eb
\eeq
being the fourier transform of the linear system's impulse response function.

From the above equations we readily find the spectral densities of $\hst_2(t)$ as:
\beq
S_{HH}(\omega) = | p(\omega-\oma) |^{-2} \left(\oma C^m_{21} A_1\right)^2\frac{4k_BT\omega_m}{MQ_m}
\eeq

\subsection{Other noise sources}
\subsubsection{Master oscillator phase noise}
\label{sec:lonoise}
To operate our device we have to feed microwave power into one resonant mode (say mode $1$), in order to detect the energy transfer between the full and the initially empty mode driven by the external perturbation.

To feed power into our device we shall use a voltage controlled microwave oscillator locked on mode $1$, at frequency $\oms$. The master oscillator phase noise is filtered through the resonator linewidth; the power spectral density has the following frequency dependence \cite{slater}:
\beq
\label{eq:powdiss}
S_{LO}(\omega) = \frac{4 \beta P_I/(\oms \qst)}{(1/\qst)^2 + (\omega/\oms - \oms/\omega)^2}
\eeq 
where $P_I$ is the power input level and $\beta$ is the coupling coefficient of mode $1$ to the output load. 
From the above equation we can estimate the microwave power noise spectral density at the detection frequency $\oma$:
\beq
\label{eq:powoma}
S_{LO}(\oma) = \frac{4 \beta P_I/(\oms \qst)}{(1/\qst)^2 + (\oma/\oms - \oms/\oma)^2} \approx \beta\frac{P_I}{\oms\qst}\left(\frac{\oma}{\oma-\oms}\right)^2
\eeq 

This figure can be improved if the receiver discriminates the parity of the e.m. field at frequency $\oma$, i.e. if it is sensitive only to the power excited in mode number $2$, rejecting all contributions coming from mode number $1$. In this way mode $1$ becomes decoupled from the output load and $\beta = 0$. The experimental set--up, based on the use of two magic--tees which accomplishes this issue is discussed in detail in \cite{rsi}. Of course the mode discrimination cannot be ideal, and some power leaking from mode $1$ to the detector's output will be present. Nevertheless our previous work has demonstrated that with a careful tuning of the detection electronics we can obtain $\beta \leq 10^{-14}$ \cite{rsi}. 

\subsubsection{Amplifier noise}
\label{sec:amplinoise}
The input Johnson noise of the first amplifier in the detection electronics has to be added to the previous contributions to establish the overall noise spectral density.
It can be described by the frequency independent spectral density \cite{papoulis}:
\beq
\label{eq:johnson}
S_{JJ} = k_B T \times 10^{(N/10)} \equiv k_B T_{eq}
\eeq
where $N$ is the noise figure of the amplifier (in dB) and $T$ the operating temperature.

\section{Detector sensitivity}
The detector sensitivity is ultimately determined by the overall effect of the various noise sources discussed in section \ref{sec:noise} and, eventually, by several others. Of course, depending on the characteristics of the system and on the experimental set--up, different noise surces will become dominant. 

We shall characterize the noise in our detector by a frequency dependent spectral density $S_n(f)$, with dimension Hz$^{-1}$, defined as follows \cite{thorne}: if a sinusoidal g.w. with known phase $\phi$, known frequency $f$ and unknown r.m.s. amplitude $\sqrt{2}h_0$, impinges on the detector, and if we try to detect the wave by fourier analyzing the detector output with a bandwidth $\Delta f$, then the {\em amplitude} signal--to--noise ratio will be:
\beq
\frac{S}{N} = \frac{h_0}{(S_n(f)\,\Delta f)^{1/2}}
\eeq
We shall also define the minimum detectable wave amplitude (at 90\% C.L.) for a periodic source with known frequency and phase as:
\beq
\label{eq:hminf}
h_{min}(f) = 1.7\,\left ( \frac{S_n(f)}{F} \right)^{1/2}
\eeq
with dimension Hz$^{-1/2}$. In the following calculations the average pattern function value $F=2/5$ has been taken \cite{maggiore}.  

Let us focus our attention on the system mentioned in section \ref{sec:spherical} based on two spherical niobium cavities working at $\oms \approx \oma \approx 2$ GHz with a stored energy in the initially excited symmetric mode of $U_1 \approx 10$ J per cell. This is  a small--scale system with an effective length of 0.1 m and a typical weigth of 5 kg. The lowest quadrupolar mechanical mode is at $\omega_m \approx 4$ kHz. In the following we shall consider an equivalent temperature of the detection electronics $T_{eq} = 30$ K.

\begin{figure}[p]
\includegraphics{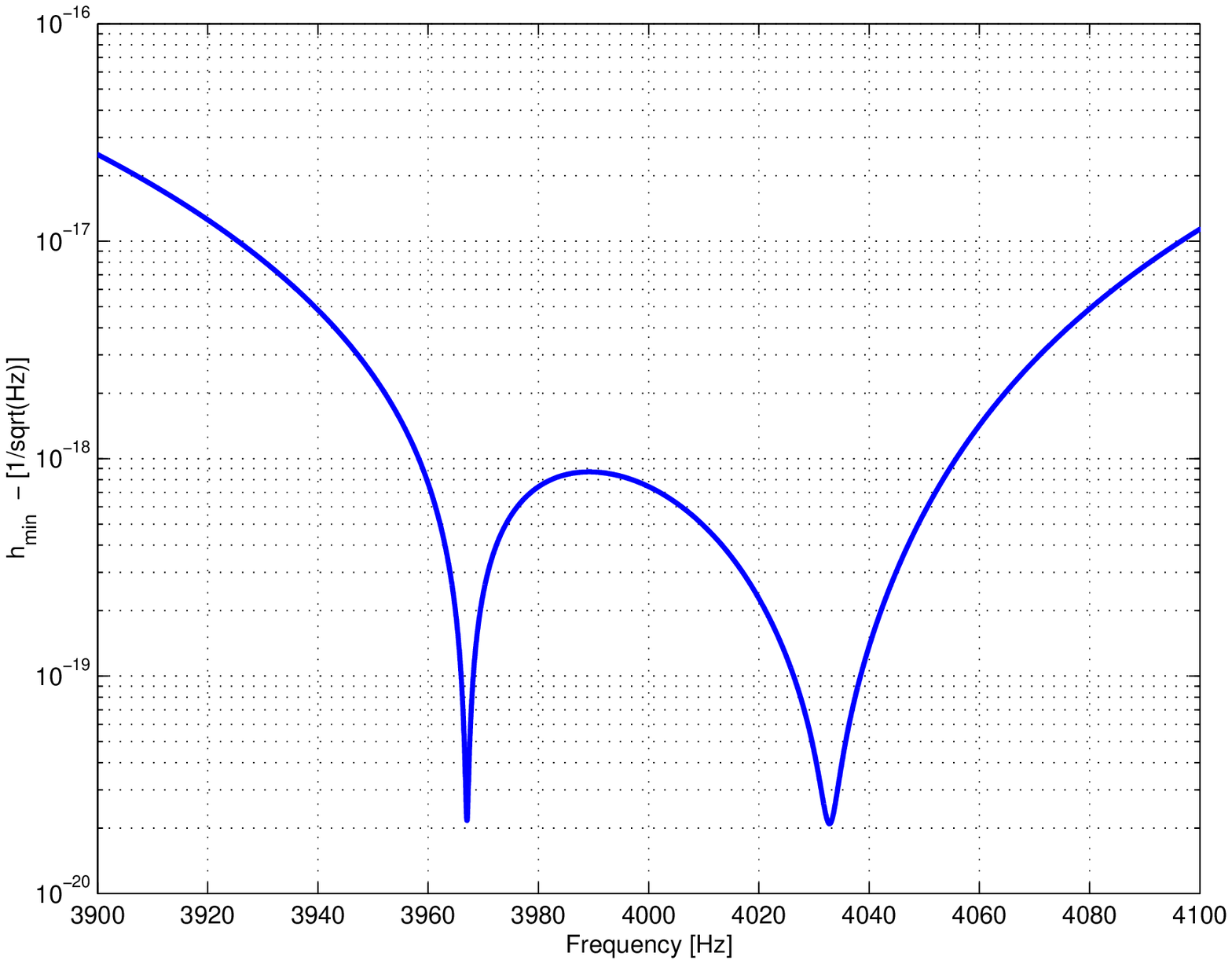}
\caption{\label{fig:hmin4} 
Calculated small--scale system sensitivity for a periodic source ($\omega_m \approx \oma - \oms = 4$ kHz, ${\mathcal{Q}}=10^{10}$, $Q_m= 10^{6}$, $T=1.8$ K, $T_{eq}=30$ K).} 
\end{figure}

\begin{figure}[p]
\includegraphics{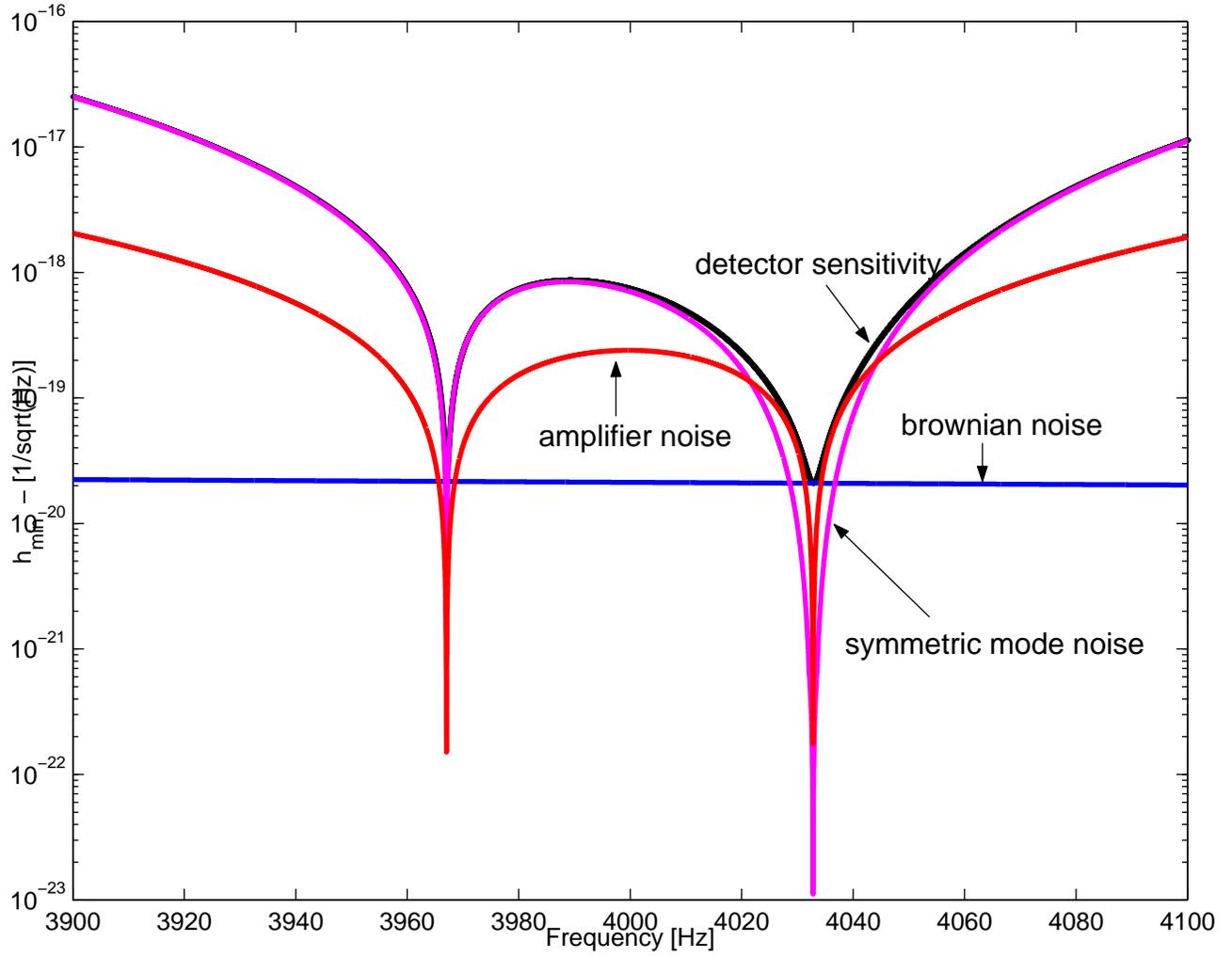}
\caption{\label{fig:vari4} 
Separate contribution of various noise sources to small--scale system sensitivity ($\omega_m \approx \oma - \oms = 4$ kHz, ${\mathcal{Q}}=10^{10}$, $Q_m= 10^{6}$, $T=1.8$ K, $T_{eq}=30$ K). As can be seen in this case the sensitivity is determined by the brownian motion of the walls while the deteciton bandwitdh is limited by the amplifier noise.} 
\end{figure}

\begin{figure}[p]
\includegraphics{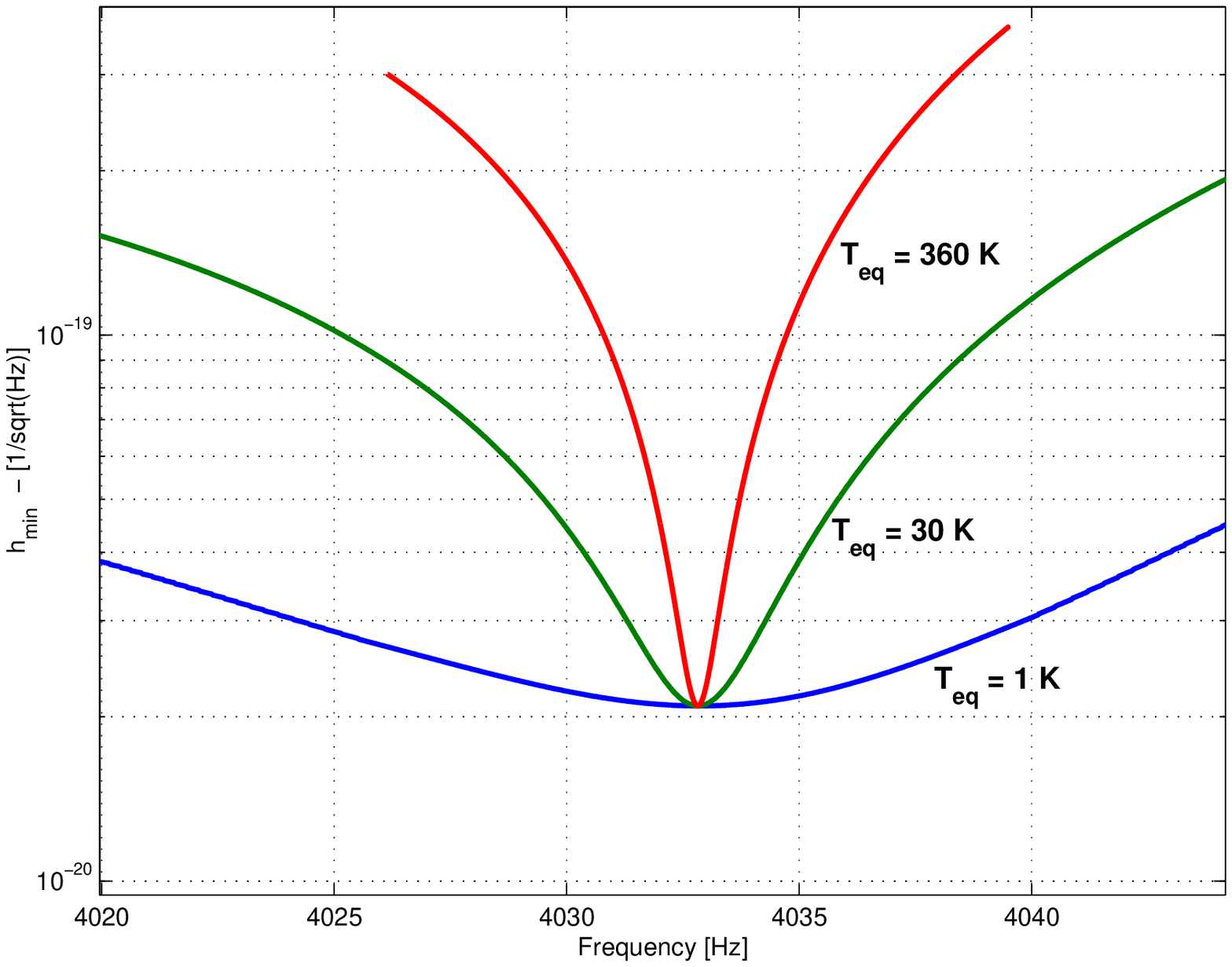}
\caption{\label{fig:bw} 
Detection bandwidth vs. $T_{eq}$ for small--scale system ($\oma - \oms = 4$ kHz, ${\mathcal{Q}}=10^{10}$, $Q_m= 10^{6}$, $T=1.8$ K).} 
\end{figure}

A possible design of the detector uses both the mechanical resonance of the structure, and the e.m. resonance. This can be accomplished if the detector is designed in 
order to have the mechanical mode frequency equal to the e.m. modes 
frequency difference $\omega_m \approx \oma - \oms$. In this frequency range, with reasonable values for the system parameters, the dominant noise source will be the noise coming from the brownian motion of the detector walls. The expected sensitivity of the detector for $\oma - \oms = \omega_m = 4$ kHz is shown in figure \ref{fig:hmin4}. In figure \ref{fig:vari4} the separate contribution of the noise sources -- discussed in section \ref{sec:noise} -- to the overall noise spectral density is shown. We point out that even if in this case the dominant noise source is the walls thermal motion a lower $T_{eq}$ would increase the detection bandwith, as shown in figure \ref{fig:bw}.

\begin{figure}[p]
\includegraphics{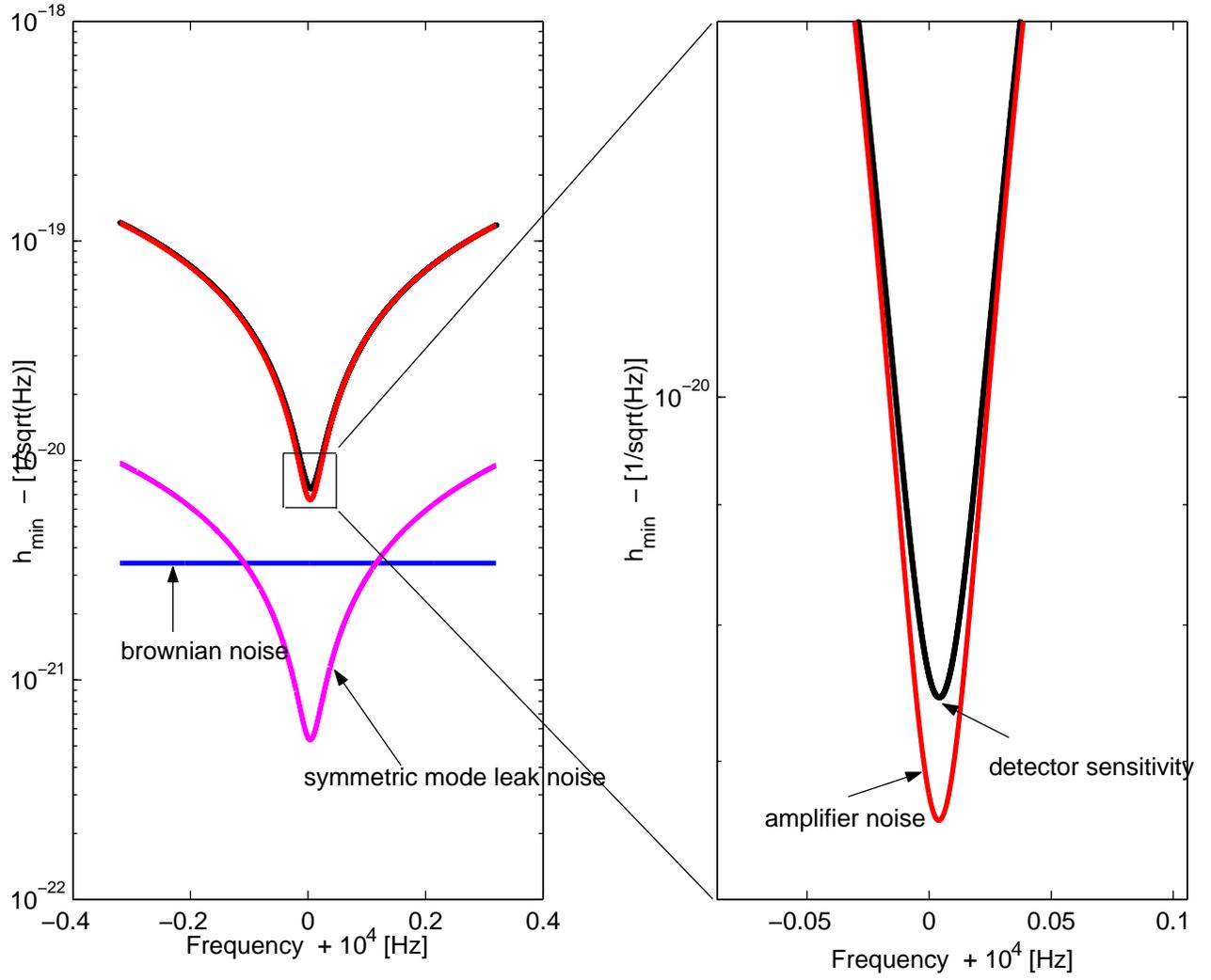}
\caption{\label{fig:vari10} 
Separate contribution of various noise sources to small--scale system sensitivity ($\omega_m = 4$ kHz, $\oma - \oms = 10$ kHz, ${\mathcal{Q}}=10^{10}$, $Q_m= 10^{6}$, $T=1.8$ K, $T_{eq}=30$ K). As can readily be seen in this case the both the sensitivity and the detection bandwidth are limited by the amplifier noise.} 
\end{figure}

\begin{figure}[p]
\includegraphics{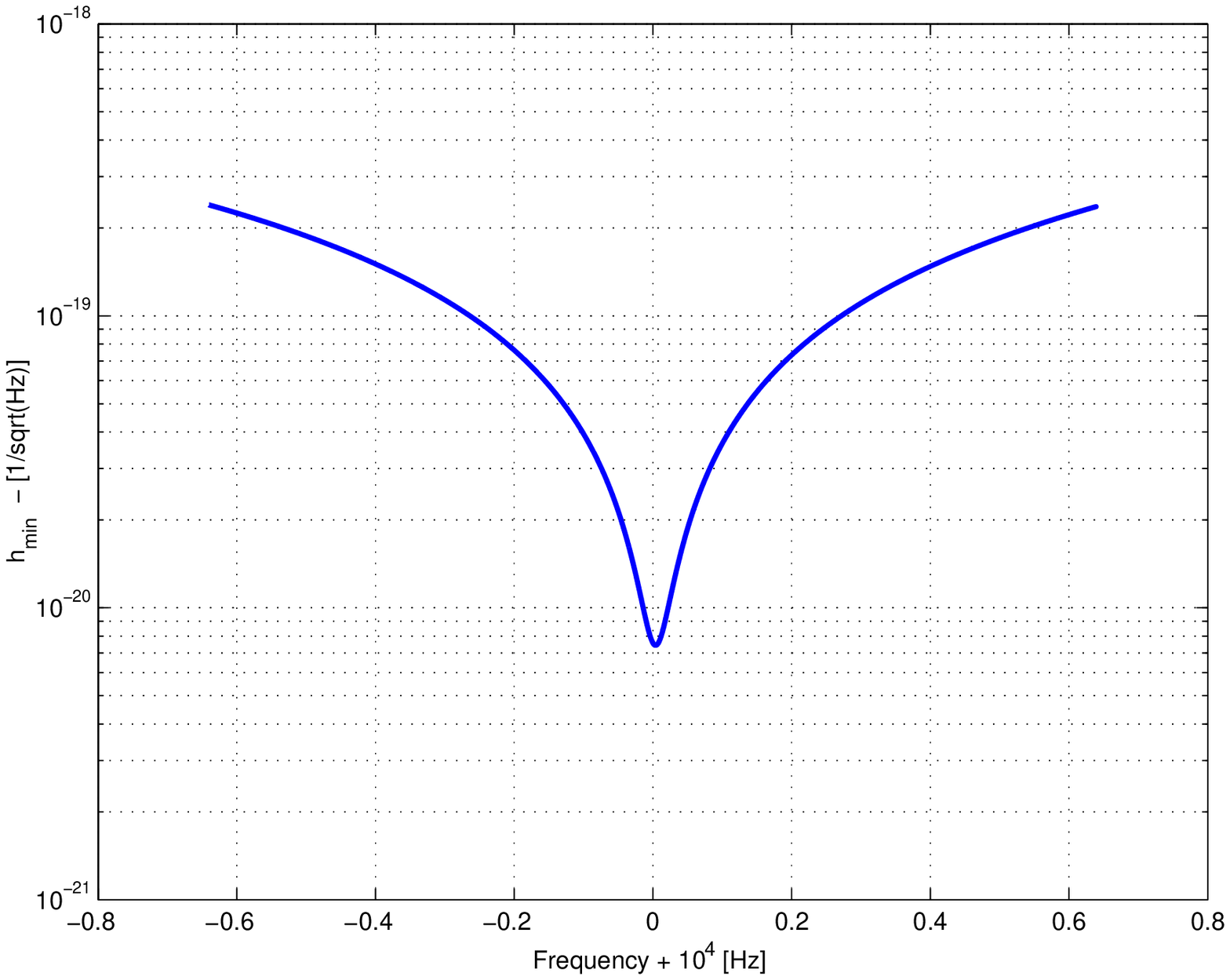}
\caption{\label{fig:hmin10} 
Calculated small--scale system sensitivity for a periodic source ($\omega_m = 4$ kHz, $\oma - \oms=10$ kHz, ${\mathcal{Q}}=10^{10}$, $Q_m= 10^{6}$, $T=1.8$ K, $T_{eq}=30$ K).} 
\end{figure}

Since our detector is based on a double resonant system (the mechanical resonator and the electromagnetic resonator) it can be operated also for frequencies $\oma-\oms \neq \omega_m$. At frequencies $\oma-\oms \leq 1$ kHz the master oscillator phase noise will, in general, be dominant (see sec. \ref{sec:lonoise}), while at frequencies $\oma-\oms \geq 10$ kHz the noise coming from the detection electronics will dominate (at least for $T_{eq}\approx 30$ K), as shown in figure \ref{fig:vari10}. The expected sensitivity of the detector for $\oma - \oms = 10$ kHz is shown in figure \ref{fig:hmin10}. 

\begin{figure}[p]
\includegraphics{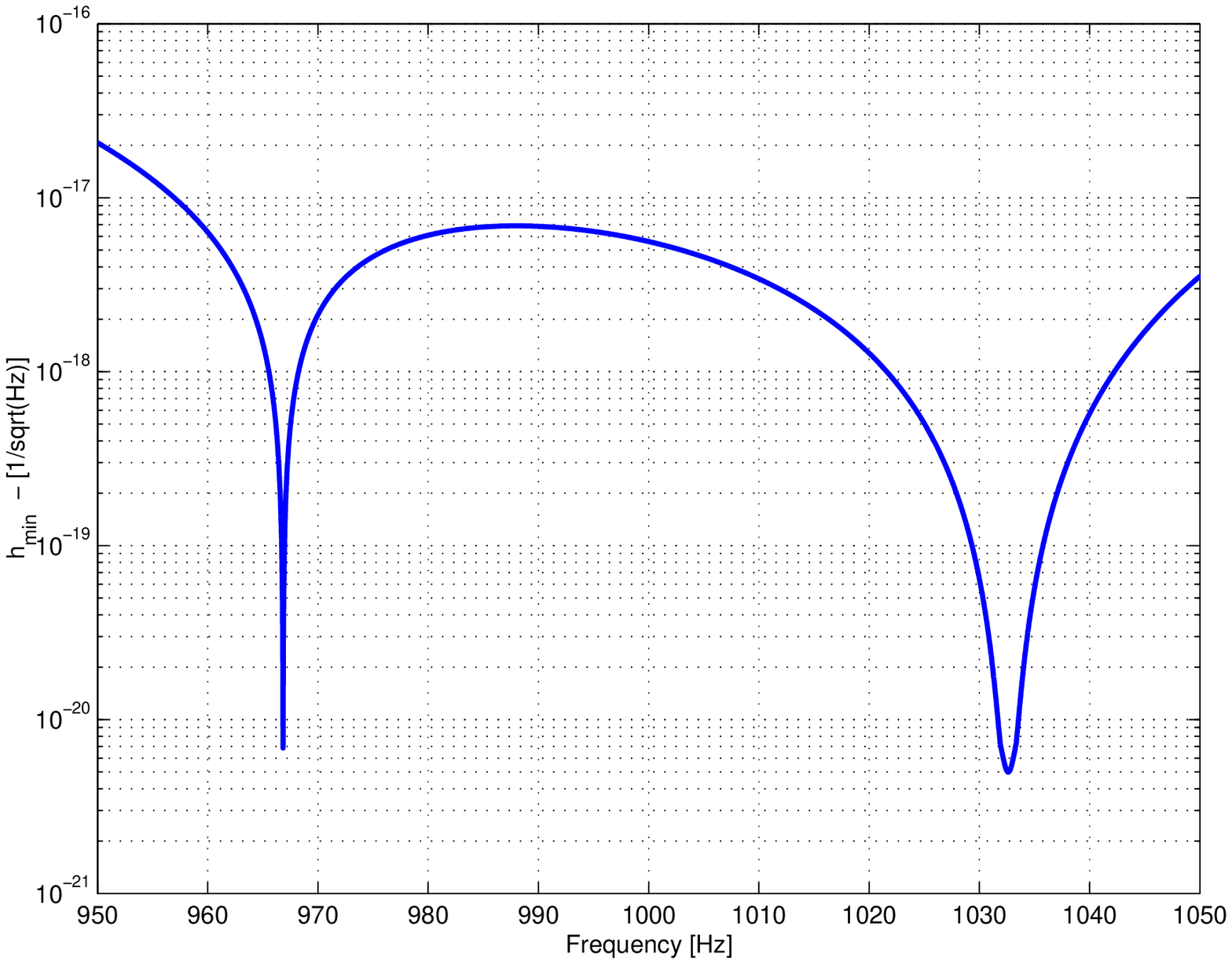}
\caption{\label{fig:hmin1} 
Calculated large--scale system sensitivity for a periodic source ($\omega_m \approx \oma - \oms \approx 1$ kHz, ${\mathcal{Q}}=10^{10}$, $Q_m= 10^{6}$, $T=1.8$ K, $T_{eq}=1$ K).} 
\end{figure}

\begin{figure}[p]
\includegraphics{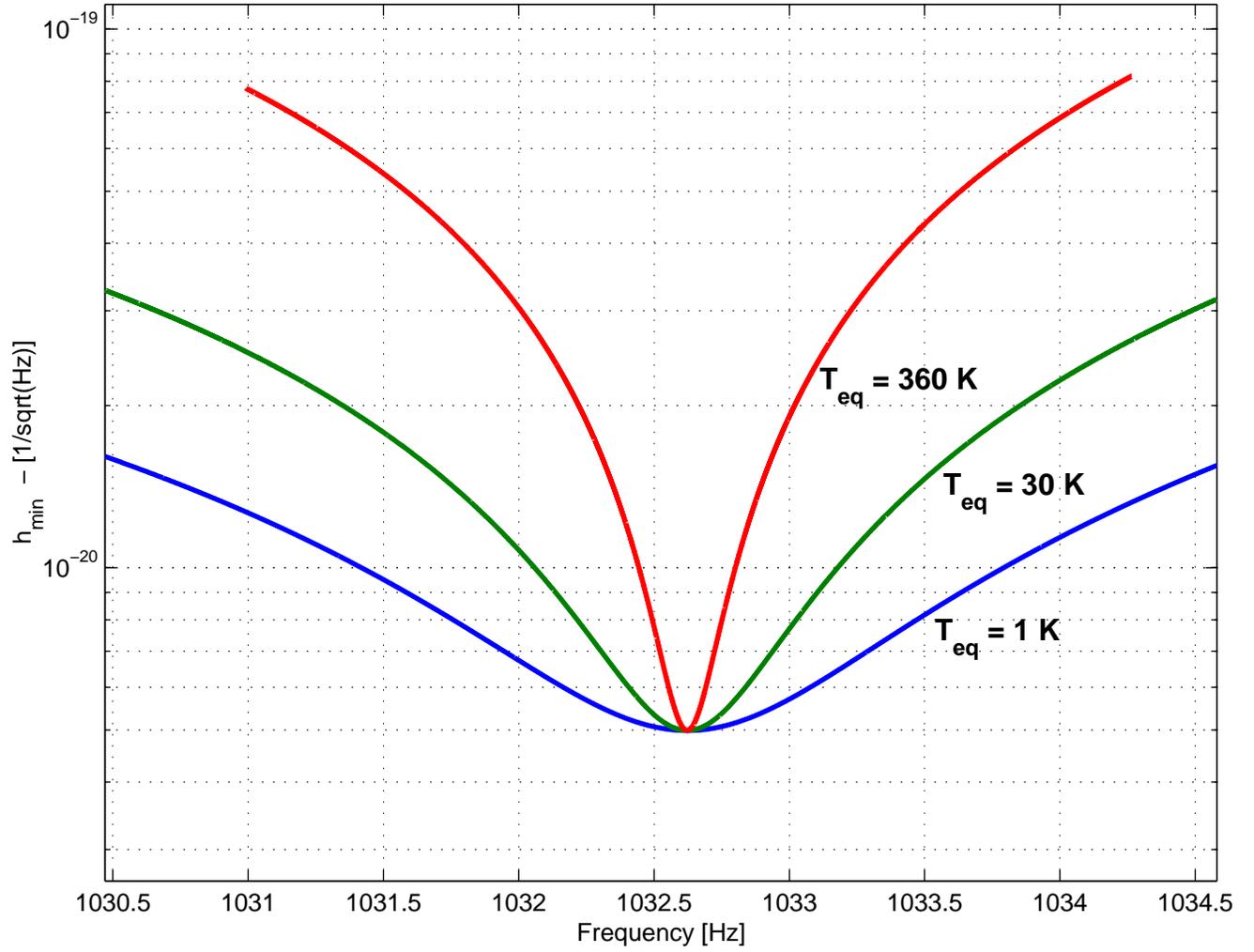}
\caption{\label{fig:bwg} 
Detection bandwidth vs. $T_{eq}$ for large--scale system ($\omega_m \approx \oma - \oms \approx 1$ kHz, ${\mathcal{Q}}=10^{10}$, $Q_m= 10^{6}$, $T=1.8$ K).} 
\end{figure}

In order to work at frequencies $\oma-\oms \leq 1$ kHz a large--scale system has to be developed. A possible design could be based on two spherical cavities working at $\oms \approx \oma \approx 500$ MHz, with $\oma-\oms \approx 1$ kHz. This system could have a stored energy of $U_1 \approx 800$ J per cell, an effective length of 0.4 m and a typical weigth of 300 kg. With a rather optimistic (but not unrealistic) choice of system parameters one could obtain the sensitivity shown in figure \ref{fig:hmin1}. We point out that in this figure an electronics equivalent temperature of $T_{eq}\approx 1$ K has been used; also in this case lowering $T_{eq}$ corresponds to an increase of the detection bandwidth (see fig. \ref{fig:bwg}).

\begin{figure}[p]
\includegraphics{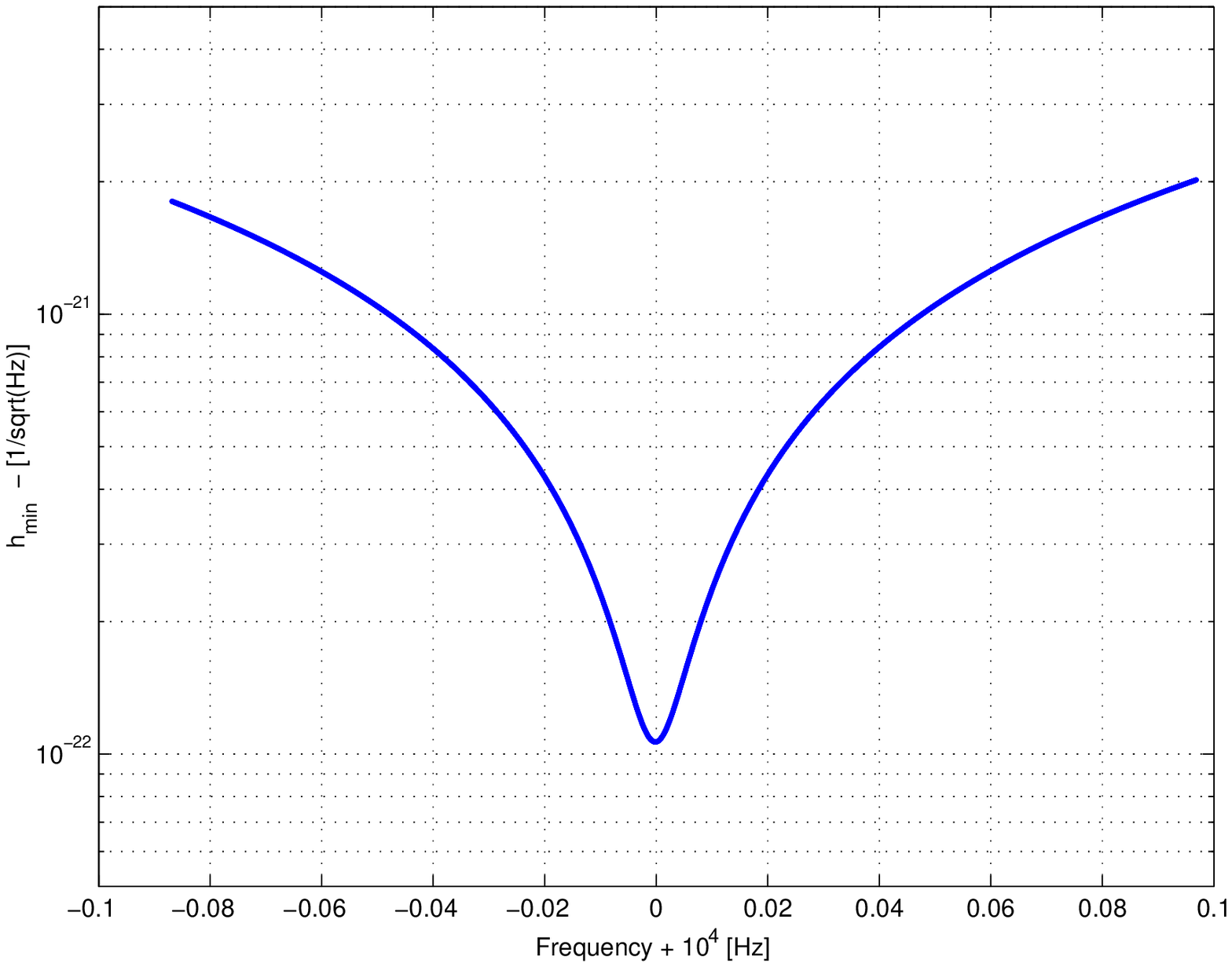}
\caption{\label{fig:hmin10g} 
Calculated large--scale system sensitivity for a periodic source ($\omega_m \approx 1$ kHz, $\oma - \oms = 10$ kHz, ${\mathcal{Q}}=10^{10}$, $Q_m= 10^{6}$, $T=1.8$ K, $T_{eq}=1$ K).} 
\end{figure}

The large--scale system could also be used at higher frequencies; in this case a good sensitivity can be achieved in a narrow detection bandwidth (see fig. \ref{fig:hmin10g}).

\section{Conclusions}
A first prototype of the detector has been built and successfully tested \cite{rsi}. A detector based on two coupled spherical cavities has been designed and preliminar mechanical and electromagnetic tests are being made on normal conducting prototypes. The planned timeline is as follows:
\begin{itemize}
\item{In 2002 a bulk niobium detector (coupled spherical cavities, $\omega = 2$ GHz, $\oma-\oms = 10$ kHz, fixed coupling) will be built at CERN;}
\item{In 2003 a variable coupling detector will be built and tested.}
\end{itemize}  

In the meantime several open problems must be addressed: 
\begin{itemize}
\item{The mechanical quality factor of the detector has to be maximized in order to suppress the noise coming from the brownian motion of the detector walls. Since mechanical dissipations arise from materials intrinsic losses and from the coupling of the system to the external environment, materials with low intrinsic losses must be used for the construction of the detector and the design of a suitable suspension system has to be done carefully.}
\item{The requirement of an high mechanical quality factor has to be matched with the requirement of high electromagnetic quality factor. This can be accomplished by the use of bulk niobium, which, at low temperatures, has low intrinsic losses both mechanical and electromagnetic, or by the use of a niobium thin film deposited on a high mechanical quality factor substrate. Both tecniques present in principle advantages and drawbacks. Several prototypes of single--cell, seamles, copper spherical cavities have been built at INFN--LNL by E. Palmieri and will be sputter--coated and tested at CERN to check the quality of noibium films deposited on spherical substrates.}
\item{A cryogenic system with a cooling power of $\approx 5$ W at $T\approx 1.8$ K and $P = 1$ bar has to be designed. The contribution of the cryogenic system to the noise has to be studied carefully.}
\item{The readout electronics has to be optimized. The use of a low noise transducer, possibly based on the SQUID technology, has to be investigated.}
\end{itemize}  

If experimental results will be encouraging, by the end of 2003 a proposal for the construction of a g.w detector, based on superconducting rf cavities could be considered.

\bibliographystyle{unsrt}
\bibliography{paco}

\begin{figure}[p]
\includegraphics{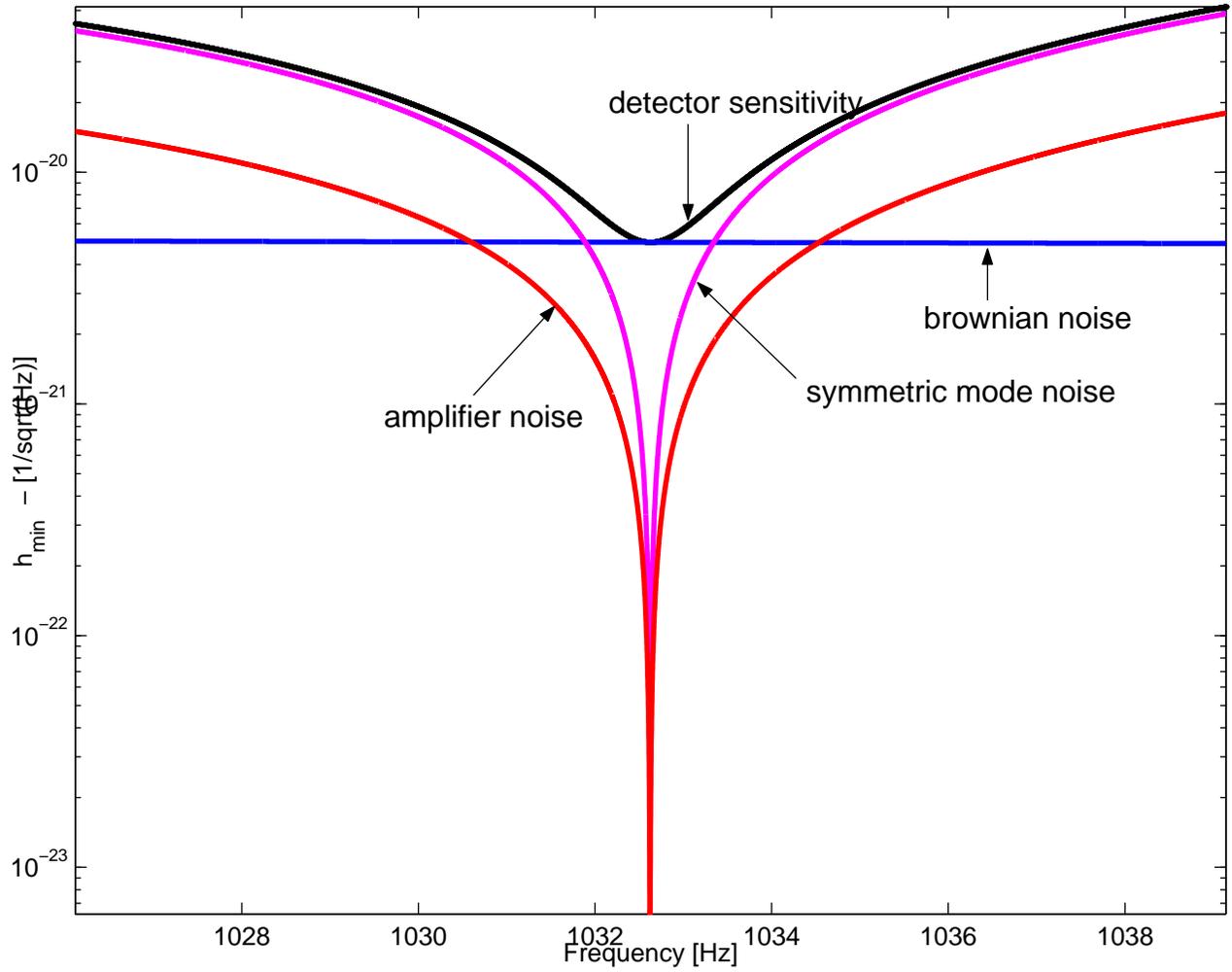}
\caption{\label{fig:vari1} 
Separate contribution of various noise sources to large--scale system sensitivity ($\omega_m \approx \oma - \oms \approx 1$ kHz, ${\mathcal{Q}}=10^{10}$, $Q_m= 10^{6}$, $T=1.8$ K, $T_{eq}=1$ K). Note that here the sensitivity is limited by the brownian noise while the detection bandwidth is set by the master oscillator phase noise.} 
\end{figure} 

\end{document}